\documentclass[11pt]{article}
\usepackage[utf8]{inputenc}
\usepackage{amsmath,amssymb,amsfonts}
\usepackage{graphicx}
\usepackage{url}
\usepackage{authblk}
\usepackage[margin=1in]{geometry}
\usepackage{natbib}
\usepackage{hyperref}
\hypersetup{
    colorlinks=true,
    linkcolor=blue,
    citecolor=blue,
    urlcolor=blue,
}
\newcommand{\keywords}[1]{\vspace{2mm}\noindent\textbf{Keywords:} #1\vspace{2mm}}
\title{An Investigation into Maintenance Support for Neural Networks}
\author[1]{Fatema Tuz Zohra}
\author[2]{Brittany Johnson}
\affil[1]{Department of Computer Science, George Mason University}
\affil[2]{Department of Computer Science, George Mason University}
\date{} 

\begin{document}

\maketitle

\begin{abstract}
As the potential for neural networks to augment our daily lives grows, ensuring their quality through effective testing, debugging, and maintenance is essential. This is especially the case as we acknowledge the prospects of negative impacts from these technologies. Traditional software engineering methods, such as testing and debugging, have proven effective in maintaining software quality; however, they reveal significant research and practice gaps in maintaining neural networks. In particular, there is a limited understanding of how practitioners currently address challenges related to understanding and mitigating undesirable behaviors in neural networks. In our ongoing research, we explore the current state of research and practice in maintaining neural networks by curating insights from practitioners through a preliminary study involving interviews and supporting survey responses. Our findings thus far indicate that existing tools primarily concentrate on building and training models. While these tools can be beneficial, they often fall short of supporting practitioners' understanding and addressing the underlying causes of unexpected model behavior. By evaluating current procedures and identifying the limitations of traditional methodologies, our study aims to offer a developer-centric perspective on where current practices fall short and highlight opportunities for improving maintenance support in neural networks.
\end{abstract}
\keywords{maintenance, debugging, testing, neural network, automated tools}

\section{Introduction and Motivation}
Neural networks have influenced various aspects of our daily lives due to their complex structure and capacity to analyze and acquire knowledge from vast quantities of data~\cite{dreyfus2005neural}. Their applications cover numerous fields, including computer vision~\cite{khan2018guide}, natural language processing~\cite{goldberg2016primer}, the health sector~\cite{shahid2019applications}, and the advancement of autonomous vehicles~\cite{chen2021deep}. This increasing adoption of neural networks in important sectors highlights their significant impact on industry as well as daily life. While these networks offer numerous advantages, even minor failures can have serious consequences, posing significant risks to human safety~\cite{Gibbs2017, penn2024algorithms}. 

The integration of a neural network into a software system requires maintaining it like any other system component.\footnote{https://aws.amazon.com/what-is/neural-network/} While software maintenance ensures the reliability and adaptability of traditional software systems, maintaining neural networks involves additional complexities. More specifically, identifying or fixing any unexpected behavior in neural networks presents unique challenges that differ significantly from traditional software due to their data-driven nature~\cite{rehman2023metamorphic}.

Consider an ML engineer, Alex. Their team relies on a deep learning model for automated document classification, and Alex is responsible for maintaining its performance over time. Alex sets up TensorBoard~\footnote{https://www.tensorflow.org/tensorboard} and Weights \& Biases (W\&B)~\footnote{https://wandb.ai/site/} to track training metrics. When model accuracy drops, they investigate using SHAP~\footnote{https://shap.readthedocs.io/en/latest/generated/shap.Explainer.html} for explainability and DeepChecks~\footnote{https://github.com/deepchecks/deepchecks} for model validation. These tools highlight potential issues—data drift, inconsistent feature importance, and unstable embeddings—but they do not provide Alex with a clear path to understanding the underlying causes. Alex spends weeks retraining, tweaking data preprocessing, and adjusting hyperparameters. They wonder: \textit{``Am I using the right tools? Are other practitioners facing the same issues? Is there a better way to systematically test, debug, and maintain neural networks?''}

Existing research on maintaining neural networks spans various approaches in providing testing techniques to assess and improve the quality of neural networks~\cite{huang2020survey, riccio2020testing}. These techniques and tools identify the presence of an issue, but they often fall short in providing an in-depth analysis of the root cause of this error or resolving it~\cite{2020survey}. Most of the existing work in localizing faults leverages activation maps, attention mechanisms, and coverage-guided fuzz testing to pinpoint faults and aid in repair processes~\cite{eniser2019deepfault, wardat2021deeplocalize, xie2019deephunter, wardat2024deepcnn}. These efforts are often tailored to specific neural network architectures, such as convolutional neural networks, limiting their general applicability~\cite{cao2022deepfd}. Static and dynamic analysis methods provide valuable insights into neural network behavior by examining model architectures and training processes. Tools that monitor gradient stability or memory usage during training help identify issues like vanishing gradients and bottlenecks~\cite{wardat2021deeplocalize, schoop2021umlaut}. Recent efforts focus on leveraging causality and neuron activation analysis to support this understanding~\cite{sun2022causality, mcqueary2024py}. 
Tools like DeepConcolic and DeepCover provide functionality for generating test cases and tracing errors to specific neurons or layers~\cite{swrhkk2018, trustai2024deepcover}, while tools like COCKPIT and DeepChecker support dynamic analysis during training, helping developers identify issues such as vanishing gradients and memory bottlenecks~\cite{cockpit2023, thedeepchecker2024}. These approaches aim to make neural network behavior more transparent, which is critical for practitioners working on finding unexpected behaviors. 
However, there is little to no work on empirical evidence regarding how effective these existing techniques and tools are in real-world practice. This raises an important question: \textit{``To what extent do practitioners adopt and utilize existing techniques and tools?''}

To fill this gap in understanding, we conducted a preliminary study involving surveys and interviews to answer the following research questions:
\begin{itemize}
    \item \textbf{RQ1:} What are the key challenges developers encounter in testing, debugging, and repairing neural networks during maintenance?
    \item \textbf{RQ2:} How effective are existing tools in supporting developers’ workflows for debugging and maintaining neural networks?
    \item \textbf{RQ3:} What features or resources do developers identify as critical for improving the maintenance tool support for neural networks?
\end{itemize}

By curating insights from practitioners, we seek to understand the practical challenges of neural network maintenance and reveal the shortcomings of current techniques and tools. Our preliminary findings aim to close the gap between research and practice, providing practitioners with the necessary tools and techniques for building reliable and trustworthy neural network systems.

\section{Methodology}
Our study aims to better understand practitioners' practices, experiences, and challenges when maintaining neural networks, particularly in testing, debugging, and repairing tasks. 
This section describes the methodology we used to conduct interviews and gather insights from practitioners. 
We began our process in accordance with ethical research standards by obtaining approval from our Institutional Review Board.~\footnote{This research is approved under IRBNet Number 2216267-1}

\subsection{Participant Recruitment and Selection}
We implemented a two-step participant selection procedure to find qualified engineers to interview. We started with an interview sign-up survey to help us identify potential engineers for interviews. Our survey asked demographic and background questions to determine participants' current employment, years of neural network experience, and knowledge of/experience with debugging traditional and AI-based systems. We posted our interview sign-up survey on LinkedIn. We also advertised within our university and professional networks via email. This initial effort yielded 228 respondents. Given that we distributed our survey online, we encountered many invalid responses. Therefore, we filtered out responses in two ways. First, we added an attention question to the interview sign-up survey. If a respondent selected the incorrect option for the attention question, we invalidated that response. This initial filtering left us with 172 responses. 

To further filter our responses to increase the likelihood of contacting quality interview candidates, we selected small subsets of respondents at random and sent a pre-assessment form. This pre-assessment included open-ended questions, inviting participants to describe specific debugging experiences, detailing the issues they encountered, the tools or methods they used, and the outcomes of their efforts. We selected potential participants based on the submitted response. We prioritized individuals based on the depth and clarity of their responses to an open question detailing practical neural network maintenance experience. For example, one participant shared how they detected and fixed exploding middle-layer representations by analyzing the min/max values of each layer and implementing normalization. Another participant mentioned their debugging methodologies, ranging from manual debugging techniques to analyzing loss curves for insights into model performance. Their detailed responses suggested the practical, hands-on experience we were looking for. This has led to four interviews so far (I1, I2, I3, and I4) with academic researchers who have practical experience maintaining neural networks. One individual (I2) informed us of having industry experience in debugging neural networks using tools, in addition to their academic research.

\subsection{Designing and Conducting Interviews}
We designed the interview script based on our research questions, ensuring it was both comprehensive and respectful of participants' experiences and insights. 
Our research artifacts are publicly available.~\footnote{https://github.com/f4zohra/Neural-Network-Maintenance-Interview.git} We begin each interview with an ice-breaking session by asking participants about their development background, experiences with neural networks, and how they integrate these tools into their workflow. 

To address RQ1, we ask participants about the types of issues they encounter, the tools or practices they use in their workflows, and the strategies they practice to resolve common issues. 
To address RQ2, we analyze the following: (1) the prevalence and reliability of debugging, testing, and repairing tools; (2) the challenges associated with the usability, documentation, and support of tools and practices; (3) alternative methods employed when tools were insufficient; and (4) experiences with customizing or comparing multiple tools. We concentrate on three key areas to answer RQ3: (1) the key features that developers consider essential for supporting neural network maintenance; (2) their interest in integration among tools; and (3) additional resources or support that could improve debugging practices. We focus on participants' personal use cases rather than task-based evaluations. Each interview lasted approximately 30–45 minutes. We recorded each interview for analysis.

\subsection{Data Analysis}

We employed a qualitative methodology to extract insights from our data after completing our interviews. We transcribed the recorded interviews and then employed the lookup summarization method to extract and correlate the data with each specific research question~\cite{braun2024thematic}. We then conducted a thematic analysis to further organize our insights. For example, when one participant stated, \textit{``There are always outliers and curveballs and exceptions to the structures you try to create. Sometimes, you get data that's in such a bizarre shape that your pipeline can't handle it,''} we extracted this quote during the lookup summarization phase and tagged it with \textbf{RQ1} and subtags for \textit{data pipeline issues} and \textit{outliers}. In the subsequent thematic analysis, we grouped this quote with other responses indicating challenges related to unpredictable data variations. This led to the formation of the theme \textbf{Data Quality Issues}, with sub-themes including \textit{Data Shape Variability}. This theme focused on the challenges developers face in maintaining neural networks.

\section{Preliminary Findings}
In this section, we will present the findings from our interviews thus far. Our goal is to explore the current state and effectiveness of neural network maintenance practices, as well as the challenges practitioners face with existing tools and techniques in real-world workflows. We seek to better understand the features and resources practitioners consider essential for supporting neural network maintenance.

\subsection{Challenges in Maintaining Neural Networks}
Prior to examining the effectiveness of current tools, it is essential to understand the inherent complexities of neural network maintenance. In this section, we will outline the issues practitioners face, setting the stage for an analysis of how these challenges impact real-world workflows.

\subsubsection{Data Quality and Preprocessing Challenges:}
One of the most common sources of failure in neural network maintenance, according to our interviews thus far, is poor input data quality and inconsistencies in preprocessing. Participants emphasized that many issues arise not from the model itself but from improperly formatted input data. For example, I1 stated, \textit{``It's not a bug, but most of the time, at the beginning, we do not get the desired output. The reason is our input. If our input is out of shape or lacks a logical pattern, the output will be just as bad.''} This highlights the importance of data quality and the need for robust preprocessing techniques. Participants also mentioned that they often have to manually check tensor shapes at different stages of the pipeline to detect inconsistencies, indicating a lack of automated techniques for this purpose.

\subsubsection{Maintaining Neural Network Gradients and Training Stability:}
According to our interviews, understanding and managing gradients during training is crucial for debugging and maintaining neural networks. Participants highlighted the limitations of traditional early-stopping techniques~\cite{prechelt2002early}. I3 noted, \textit{``People used to use early stopping when the loss is not changing much, but this is not helpful. As long as gradients are changing, the network is still learning something.''} Participants mentioned encountering issues with extremely large weight values that can lead to unexpected failures due to exceeding numerical limits. To address these challenges, practitioners often use hyperparameter tuning to adjust learning rates, weight initialization, and regularization techniques.

\subsubsection{Fairness and Bias Issues in Model Predictions:}
Ensuring fairness and mitigating biases in model predictions emerged as a key challenge in our interviews. Participants described encountering situations where models exhibited biases, particularly in tasks involving sensitive attributes like ethnicity or gender. For instance, I2 shared an experience: \textit{``The first time, the results showed that accuracy was especially poor at identifying names that sounded like they were from a certain ethnicity. So, I went out and found additional datasets with a high proportion of those names and retrained the model.''} This underscores the importance of proactive bias detection and mitigation strategies. Participants also mentioned employing techniques like data augmentation and fairness testing to ensure ongoing fairness and reliability.

\subsection{Effectiveness of Existing Maintenance Tools}
Practitioners use various tools for debugging and maintaining neural networks, but they encounter challenges related to workflow integration, usability, and a lack of automated failure analysis. Our interviews highlight that while visualization and logging tools aid in testing and debugging, many still rely on manual inspection due to the lack of comprehensive solutions.

\subsubsection{Existing Debugging and Monitoring Tools:}
Our interviews revealed that practitioners utilize a combination of manual debugging techniques and tools to address unexpected outputs and inspect intermediate representations in neural networks. 
Participants mentioned relying on tools such as Comet ML~\footnote{https://www.comet.com/site/}, Weights \& Biases~\footnote {https://pytorch.org/} and Checklist to track model behavior, visualize errors, and debug. For some participants, debugging relied less on automated tooling and more on print statements, structured data logs, and small-scale testing. One of the participants, I2, designed custom scripts to identify and analyze patterns in model failures. I2 stated, \textit{``I have some scripts that I tend to use that will, anytime a failure occurs, output details about it into a structured table.''} For I4, although they were aware of available tooling, they preferred to \textit{``ask ChatGPT to write code''} for them and then refine the output through iterative prompts, using this process to quickly prototype solutions or investigate issues.

\subsubsection{Tool Integration and Usability Challenges:}
 
  Our interviews revealed that while debugging tools offer valuable insights, their integration into developer workflows presents some challenges. According to the participants, flexibility emerged as a key factor in tool selection. Participants expressed a preference for tools that allow for customization and personalized visualizations. For example, I4 favored Comet ML over Weights \& Biases due to its ability to create custom visualizations using JavaScript, while highlighting Weights \& Biases for its predefined options. Another challenge identified by participants is the complexity mentioned in the tool documentation. I2 mentioned that in the documentation of Checklist, \textit{``so many abstractions to do so many things that there's a lot of tool-specific jargon you need to know.''}. I2 also noted: \textit{``The tutorials weren't as clear as they could have been. I had to take a lot of notes before I fully understood how to use it.''} There are also regional accessibility restrictions due to international sanctions. I4 stated, \textit{``One of the biggest [limitations] is that I’m from Iran, so Weights \& Biases is not available in Europe because of the sanctions.''}

\subsection{Improving Maintenance Support}

Researchers highlighted several missing features in current debugging tools and suggested key areas for improvement.

\subsubsection{Automated and Personalized Debugging Assistance:}
Our interviews highlighted a significant gap in current tooling regarding proactive debugging recommendations. Participants expressed a need for tools that could analyze training behavior and offer specific suggestions for addressing common issues such as overfitting and underfitting. I4 stated, \textit{``If my test loss keeps increasing, the tool should automatically tell me, ‘Try decreasing the learning rate, increasing layers, or adding more data.''} Participants also explored the potential of LLMs in neural network maintenance, suggesting personalized LLMs that could learn from their debugging habits and offer tailored suggestions. One participant suggested, \textit{``Maybe in the future, we'll have personalized LLMs that adapt to our debugging patterns.''} However,  participants also acknowledged that while LLMs can assist with code generation and pattern identification, human intuition remains crucial for debugging complex and unexpected failures.

\subsubsection{Better Learning Resources:}
A major concern raised by participants was the lack of comprehensive learning resources specifically focused on debugging neural networks. They noted that most online courses and tutorials primarily teach how to build models, neglecting the crucial aspect of debugging. As I3 stated, \textit{``Every tutorial teaches you how to build a model, but none explain what might go wrong and how to debug it.''} Participants also highlighted the reliance on clean, preprocessed 'toy datasets' in tutorials, arguing that these datasets fail to reflect the complexities of real-world data. This discrepancy between academic training and industry practice creates difficulties for practitioners when faced with noisy, real-world datasets. I3 emphasized this gap, stating, \textit{``When you are dealing with a new dataset that you collected in the real world, it is really ugly.''} Participants emphasized the need for practical, comprehensive learning resources that address the challenges of debugging and maintaining models using real-world data.

\subsubsection{Human-in-the-Loop Support:}

Our interviews revealed skepticism among practitioners regarding fully automated debugging tools. While acknowledging the potential of automation for certain tasks, participants emphasized the importance of human intuition and expertise in navigating complex debugging scenarios. I1 stated, \textit{``If debugging is just about checking tensor shapes, a tool can automate it. But deeper issues require human intervention. An integrated tool that does everything—testing, debugging, and repairing—sounds nice, but in reality, it’s hard to maintain as datasets and domains evolve.''} Instead of a unified, fully automated approach, participants favored a more modular approach, with tools that can be adapted to different tasks and integrated into human-driven workflows. I2 expressed concern about the assumptions made by integrated tools, stating, \textit{``The more things that an integrated tool tries to do at once, the more assumptions it has to make, which means the sooner one of those assumptions gets broken in a way that is really not good for the long-term usefulness of the tool.''}

\section{Discussion}
Our initial findings highlight several key insights into improving tools and practices for debugging neural networks, which we aim to build on in our ongoing work. Given the growing ubiquity and complexity of neural networks as solutions, it is vital that we are able to provide engineers with tooling that goes beyond the identification of problems into support for understanding and thus effectively repairing them.  This also emphasizes the need to better understand existing support for repairing defects in neural networks and how tools that aim to provide an understanding of model behavior, like Pyholmes~\cite{inspired2024pyholmes}, can best support the repair and validation process.

Our findings also indicate a potential misalignment between academic training and preparation for real-world debugging scenarios. Our participants, especially I3 and I4, pointed to the prevalence of "toy datasets" in tutorials, which fail to reflect the complexities of real-world data and leave engineers underprepared for real-world debugging scenarios. The lack of formal education in testing and debugging practices can lead engineers to adopt ad-hoc solutions, potentially resulting in suboptimal or inefficient maintenance processes for neural networks. This realization has led to the creation and evolution of a curriculum centered on aligning AI engineering with traditional software engineering foundations that support maintainability~\cite{kastner2020teaching}. It also points to a need for more practical educational resources that address real-world debugging challenges, incorporating diverse datasets and common issues faced by practitioners. 

Our preliminary findings also suggest that AI-assisted tools could play a crucial role in streamlining the debugging process and helping developers locate issues more efficiently. The increasing use of large language models (LLMs) such as ChatGPT for generating initial code and debugging suggestions demonstrates the growing role of AI-assisted development. Our participants conveyed how important human intuition and experience are in complex debugging situations, which suggests that a human-in-the-loop approach, in which AI tools help human judgment instead of replacing it, may work better.  In the future, one of the biggest research challenges will be to determine if and to what extent AI-assisted tools that work across domains can still provide useful, case-specific debugging information. This will require careful consideration of how to balance the benefits of automation with the need for human oversight and expertise.

\section{Conclusion \& Future Work}
In this paper, we present preliminary insights into existing support for maintaining neural networks. Based on an exhaustive literature review and interviews with four academic researchers so far, we found that numerous existing efforts have emerged in recent years to support training and testing issues in neural networks. While our findings provide useful insights into the landscape of available support, they also amplifies gaps in our progress toward realizing effective maintenance support in practice. 
Our findings thus far further motivate continued efforts to understand experiences with automated support for identifying, understanding, and repairing issues in neural network behavior. While our initial study involved a small sample size, we will continue to broaden participation by including more industry practitioners and engineers working with diverse neural network architectures to gain a more comprehensive understanding of maintenance challenges across various domains.

\bibliographystyle{plainnat}
\bibliography{main}

\end{document}